\documentclass[aps,twocolumn,prl,showpacs]{revtex4}
\usepackage{graphicx}
\usepackage{dcolumn}
\usepackage{bm}

\newcommand{\nt}{$\nu_T=1$}
\newcommand{\dl}{$d/\ell$}
\newcommand{\dlc}{$(d/\ell)_c$}

\begin{document}

\title{Evidence for a finite temperature phase transition in a bilayer quantum Hall system}

\author{A.~R. Champagne$^1$, J.~P. Eisenstein$^1$, L.~N. Pfeiffer$^2$, and K.~W. West$^2$}

\affiliation{$^1$Condensed Matter Physics, California Institute of Technology, Pasadena CA 91125
\\
$^2$Bell Laboratories, Alcatel-Lucent, Murray Hill, NJ 07974}

\date{\today}

\begin{abstract}
We study the Josephson-like interlayer tunneling signature of the strongly correlated \nt\ quantum Hall phase in bilayer two-dimensional electron systems as a function of the layer separation, temperature and interlayer charge imbalance. Our results offer strong evidence that a finite temperature phase transition separates the interlayer coherent phase from incoherent phases which lack strong interlayer correlations. The transition temperature is dependent on both the layer spacing and charge imbalance between the layers.
\end{abstract}


\maketitle

Bilayer two-dimensional electron systems (2DES) at high magnetic fields can exhibit drastically different quantum collective phases depending on whether their interlayer spacing is large or small. When the spacing is large the two layers act independently and display the familiar fractional quantum Hall and related effects. Conversely, at small interlayer separation, bilayer collective phases with no single layer analog appear \cite{perspectives}.

An especially interesting example of this occurs when the total density $n_T$ of electrons in the bilayer equals the degeneracy $eB/h$ of a single spin-resolved Landau level created by the magnetic field $B$.  In this situation the total Landau level filling factor is $\nu_T = n_T/(eB/h) = 1$. If the spacing between the two layers is small, the 2DES is a gapped quantum Hall effect (QHE) liquid\cite{perspectives} with several very unusual properties, including Josephson-like interlayer tunneling\cite{Spielman00} and vanishing Hall and
longitudinal resistances\cite{Kellogg04,Tutuc04,Wiersma04} when currents are driven in opposition (counterflow) in the two layers.  For layer spacings larger than a critical value the system properties revert to those characteristic of independent layers. Interlayer tunneling is heavily suppressed, no anomalous counterflow transport properties are observed and, for equal density layers (i.e. with individual filling factors $\nu_{top} = \nu_{bot} = 1/2$), there is no quantized Hall effect.

There now exists a large theoretical literature dealing with the strongly correlated bilayer \nt\ QHE phase at small layer separation.  It is widely believed that the system is well described as an easy-plane ferromagnet
with the layer index (``top'' or ``bottom'') of the electrons regarded as a pseudospin quantum number (``up'' or ``down''). Exchange interactions favor a configuration in which all electrons occupy a single coherent linear
combination of up and down pseudospin states.  Interlayer charging effects favor equal amplitudes of the two states and thus the net pseudospin moment lies near the $x-y$ plane.  In the limit of zero tunneling the transition to this coherent state is believed to be spontaneous. At the qualitative level this picture accounts well for many of the most dramatic experimental observations, including the existence of the QHE \cite{perspectives} in weakly tunneling samples, the strong many-body enhancement of the tunneling at zero bias \cite{Spielman00}, the presence of a linearly dispersing collective mode \cite{Spielman01}, and the peculiar counterflow transport properties \cite{Kellogg04,Tutuc04,Wiersma04}.  Quantitatively, the situation is less impressive.  For example, neither the persistence of the zero bias tunneling peak to significant in-plane magnetic fields \cite{Spielman01} nor the unexpected temperature dependence of the resistivity in counterflow \cite{Kellogg04,Tutuc04,Wiersma04} are understood.

Both experiment and theory strongly suggest that the coherent QHE phase at small layer spacing and the incoherent phase at large spacing are separated by a phase transition.  Nevertheless, the nature of that transition remains unclear.  Early theoretical work invoked a continuous quantum phase transition to the
ferromagnetic state at zero temperature and a Kosterlitz-Thouless transition at finite temperatures \cite{Wen92,Yang94}. However, the possibility that the phase transition is in fact weakly first order cannot be ruled out, and some numerical evidence for this has been reported \cite{Schliemann01}.  While experiments
typically show a continuous, if rapid, transition between the two phases, it is possible that disorder (e.g. static density fluctuations) might smooth out weakly discontinuous observables via phase separation near the critical point.  This scenario has been invoked \cite{Stern02} to explain recent Coulomb drag experiments \cite{Kellogg03} and is also consistent with recent spin polarization measurements \cite{Spielman05,Kumada05}. That the drag results have also been found to be consistent with a continuous transition between the two phases \cite{Simon03} highlights the uncertainty over the nature of the critical point.

Here we report the results of interlayer tunneling experiments which shed new light on the nature of the phase transition at \nt. We demonstrate that the dependence of the zero bias interlayer tunneling peak on layer spacing in the coherent phase allows for an accurate determination of the critical layer spacing. Furthermore, we find that while the critical layer spacing evolves smoothly with temperature, the basic dependence of the tunneling on layer spacing follows a single simple empirical formula, independent of temperature over a wide range.  These observations constitute strong evidence that a temperature-dependent phase transition separates the coherent and incoherent phases at \nt. We buttress this evidence with additional tunneling data taken at \nt\ but with unequal densities in the two layers.

\begin{figure}
\includegraphics[width=3.35in,bb= 149 86 381 248]{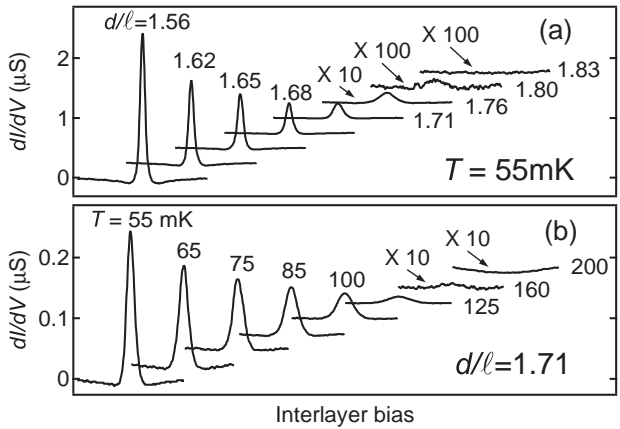}
\caption{\label{} Tunneling conductance spectra $dI/dV$ vs. $V$ at \nt. Each trace ranges from -100 $\mu$V $< V <$ +100 $\mu$V; origins are displaced for clarity. (a) Dependence on effective interlayer separation \dl\ at a
fixed low temperature of 55 mK. (b) Dependence on temperature at fixed \dl\ = 1.71.} \end{figure}

The sample used in this experiment contains two 18 nm GaAs quantum wells separated by a 10 nm Al$_{0.9}$Ga$_{0.1}$As barrier.  Remote Si doping layers populate the ground subband of each quantum well with a 2DES with nominal density $5.5 \times 10^{10}$ cm$^{-2}$ and low temperature mobility $1 \times 10^6$ cm$^2$/Vs. The active region of the sample is a square 250 $\mu$m on a side.  Independent electrical contacts to the individual layers allow measurement of the interlayer tunneling conductance $dI/dV$ versus interlayer voltage $V$. The electron density in each layer is controllable via electrostatic gating and this allows study of the $\nu_T = \nu_{top} + \nu_{bot} = 1$ state at different effective layer separations $d/\ell$ (with $d$ = 28 nm
the center-to-center quantum well separation and $\ell = (\hbar/eB)^{1/2}$ the magnetic length at \nt) in both density balanced ($\nu_{top} = \nu_{bot}$) and imbalanced ($\nu_{top} \neq \nu_{bot}$) configurations.

Figure 1 illustrates how the zero bias tunneling peak develops as the coherent \nt\ phase is entered.  In Fig. 1(a) tunneling conductance spectra ($dI/dV$ vs. $V$) taken at $T = 55$ mK and various \dl\ are shown.  At high \dl\ the tunneling conductance at zero bias is very small, being strongly suppressed by Coulomb blockade-like effects characteristic of single 2DES layers at high magnetic field \cite{Eisenstein92}. As \dl\ is reduced below a critical value (about 1.82 for the data shown) a sharply resonant enhancement of the tunneling conductance appears at zero bias.  This peak grows rapidly as \dl\ is further reduced and soon completely dominates the tunneling spectrum.  This phenomenon \cite{Spielman00} has been interpreted as signaling a quantum phase transition in the bilayer 2DES to a coherent state in which many-body effects dominate. To within experimental accuracy the onset of this peak coincides with the onset of the in-plane transport features associated with the quantized Hall effect in the same sample.  We stress, however, that tunneling is a much more sensitive probe of the onset of the coherent state than bulk transport. Not only is the tunneling feature sharp and readily distinguished from background effects, but unlike transport tunneling is a local probe which does not require percolative transport pathways for its detection.

The regime of strong zero bias tunneling can also be entered by lowering $T$ at fixed \dl, provided the latter is small enough.  Figure 1(b) shows a series of tunneling spectra taken at
$d/\ell = 1.71$. At about $T = 160$ mK a small zero bias peak becomes detectable and proceeds to grow rapidly as the temperature is reduced further.  The same basic behavior is observed for all \dl\ less than about 1.82. We find that at each effective layer separation there is a critical temperature above which the tunneling peak is either absent or unobservably small.

The basic phenomenology of the zero bias tunneling peak at \nt\ is summarized in Fig. 2(a) where we plot the zero bias tunneling conductance $G(0)$ (i.e. $dI/dV$ at $V=0$) versus \dl\ for several different temperatures, $T$ \cite{offset}. The shape of the $G(0)$ vs. \dl\ curves is qualitatively the same at all $T$. The peak conductance rapidly collapses, by up to four orders of magnitude, as \dl\ increases and extrapolates to zero at a temperature-dependent critical effective layer separation, \dlc.  Interestingly, the data show no evidence of thermal smearing in the vicinity of \dlc; the same rapid rise of the tunneling conductance below the critical layer separation is seen at all temperatures studied.

To obtain a more quantitative comparison of the data at different temperatures we use an empirical fitting procedure to extract the critical layer separation. The solid lines in Fig. 2(a) are weighted
power law fits of the form, $G(0) = K[(d/\ell)_c - (d/\ell)]^p$. Figure 2(b) shows that the empirical exponent $p$ is temperature independent at $p \approx 2.85$. The prefactor $K$ changes by less than a factor of 2 between 55 mK and 250 mK. Fig. 2(c) demonstrates that \dlc\ falls linearly with increasing temperature.

We stress that the power law fits just described are empirical and not intended to represent scaling laws in the usual sense of continuous phase transitions.  Nevertheless, these fits allow us to make two new and important conclusions about tunneling at \nt.  First, the temperature independence of the fitted exponent $p$ makes definite our qualitative observation that the tunneling conductance in the coherent \nt\ phase has a universal dependence on effective layer separation.  Second, the fits allow a consistent way to extract the critical layer separation \dlc\ and have thereby revealed the linear temperature dependence of this important parameter.

\begin{figure}
\includegraphics[width=3.35in,bb= 138 110 377 417]{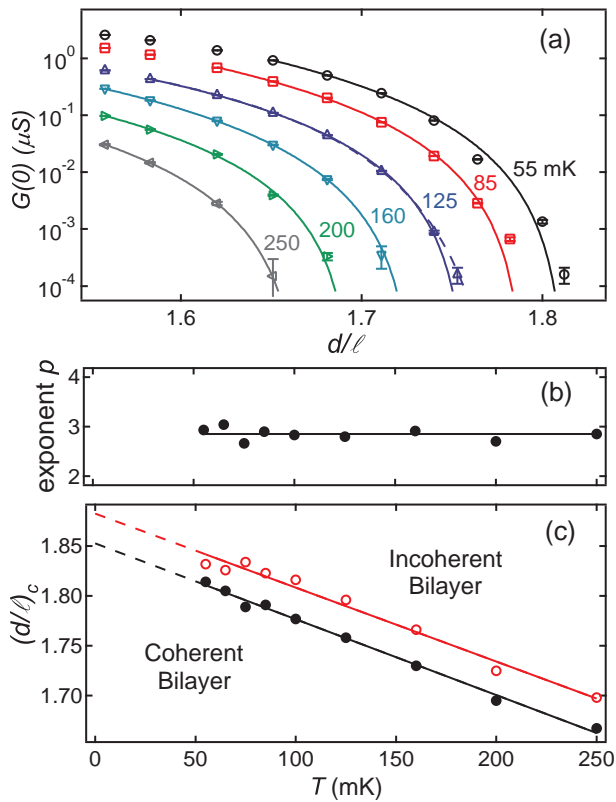}
\caption{\label{}(Color online.) (a) The conductance peak $G(0)$ at zero interlayer bias versus $d/\ell$ for various $T$. The solid lines are fits to the power law function $G(0) = K[(d/\ell)_c - (d/\ell)]^p$.  The dashed line at $T = 125$ mK is a representative fit using the exponential function $G(0)=A e^{-D/((d/\ell)_c-d/\ell)^{1/2}}$.  (b) The exponents $p$ obtained from the power law fits to the data shown in (a) and other similar data. (c) Phase boundary for the $\nu_{T}=1$ QH state. The solid circles are the critical $(d/\ell)_c$ obtained from the power law fits while the open circles are obtained from the exponential fitting function. The solid lines are linear fits to the phase boundary.}
\end{figure}
The relatively large fitted exponent $p \approx 2.85$ reflects the fact that the tunneling conductance rises quite smoothly as \dl\ is reduced below \dlc.  Indeed, we have found it possible to obtain equally good fits to the data in Fig. 2(a) using a function \cite{KTfunc} which is singularly smooth at the transition: $G(0)=A e^{-D/((d/\ell)_c-d/\ell)^{1/2}}$. For these fits, one of which is shown with a dashed line in Fig. 2(a), we find $D \approx 2.75 \pm 0.25$, essentially independent of temperature.  As before, the critical effective layer separation \dlc\ falls linearly with increasing temperature although, as Fig. 2(c) shows, the precise values are slightly larger than found with the power law fits. Thus, this fitting function leads to the same basic conclusions as the power law function.

The absence of obvious thermal smearing of the transition combined with the universal dependence of the tunneling conductance on \dl\ and the steady shift of the critical point \dlc\ with temperature all suggest that the bilayer 2DES at \nt\ undergoes a true finite temperature phase transition with Fig. 2(c) illustrating a linear relationship between the critical layer separation \dlc\ and the critical temperature $T_c$.

Additional evidence in support of a finite temperature phase transition is offered by the dependence of the tunneling conductance $G(0)$ on layer density imbalance.  Figure 3(a) displays $G(0)$ at $\nu_T =\nu_{top}+\nu_{bot}=1$ and $T = 55$ mK as a function of the filling factor difference $\Delta \nu =\nu_{top}-\nu_{bot}$ between the layers.  For $d/\ell \lesssim 1.71$ $G(0)$ exhibits a maximum at balance ($\nu_{top}=\nu_{bot}=1/2$) and drops gently as imbalance is imposed. As \dl\ is increased, the local maximum of $G(0)$ at balance is replaced by a minimum. Eventually this mininum falls to zero at the (previously defined) critical point \dlc\ and, for a narrow range of $d/\ell > (d/\ell)_c$, $G(0)$ remains zero out to a finite imbalances.  This observation strongly suggests that the phase boundary separating the coherent and incoherent
phases at \nt\ moves to larger layer separation as imbalance is imposed, a fact already appreciated \cite{Spielman04}.  What is new here is that, as Fig. 3(b) demonstrates, this phenomenology repeats itself at higher temperatures, only at larger values of \dl.  This accurate repetition of the imbalance dependence of
$G(0)$ near the phase boundary as the temperature is raised generalizes our earlier observations  on the \dl\ dependence of tunneling at balance. We suggest that the linear relation between temperature
and \dl\ shown in Fig. 2(c) is but a 1D cut through a 2D surface which defines the \nt\ phase transition in $d/\ell-T - \Delta \nu$ space.  

\begin{figure}
\includegraphics[width=3.35in, bb=68 68 386 213]{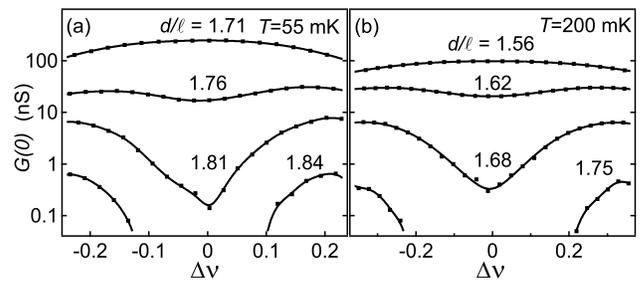}
\caption{\label{}$G(0)$ at \nt\ vs. $\Delta \nu$ in imbalanced bilayers at \nt\ for various \dl\ at $T$ = (a) 55 mK, (b) 200 mK. The solid lines are guides to the eye.}
\end{figure}
In the most developed theoretical scenario, the low energy dynamics of a bilayer 2DES at \nt\ is approximated by a modified 2D-XY model\cite{perspectives}. In this scenario the relationship\cite{Sondhi} between a $d$-dimensional quantum system at zero temperature and a classical $(d+1)$-dimensional system at a finite (pseudo-) temperature is exploited in order to assert that the \nt\ system undergoes a quantum phase transition at $T=0$ as the effective layer separation \dl\ is reduced below a critical value. The ordered state is
ferromagnetic, with all electrons occupying a specific equally weighted linear combination of the individual layer eigenstates. In the absence of tunneling this transition is spontaneous.

At finite temperatures a spontaneous transition to a ferromagnetically ordered phase is no longer possible. Instead, the system is anticipated to undergo a Kosterlitz-Thouless transition to a phase in which correlation functions decay algebraically with distance, thus lacking long range order. The KT transition temperature $T_{KT}$ is determined by the pseudospin stiffness $\rho_s$ at the transition temperature: $T_{KT} = (\pi/2) \rho_s(T_{KT})$.  A rough estimate of $T_{KT}$ follows from Hartree-Fock calculations of $\rho_s$ at $T=0$ where, in limit of very small layer separation, $\rho_{s,0} = 0.0249 e^2/\epsilon\ell$ \cite{Moon95}. The spin stiffness falls as \dl\ increases, with quantum fluctuations forcing it to vanish altogether at a zero temperature critical layer separation estimated to be about $(d/\ell)_{c,0} \sim 1.8$ for the parameters of the current samples \cite{Schliemann01}. If, for simplicity, we assume that $T_{KT}$ falls linearly from $(\pi/2)\rho_{s,0}$ to zero as \dl\ increases from zero to 1.8, we find $dT_{KT}/d(d/\ell) \approx -1.7 \pm 0.1$ K over the range of \dl\ relevant here.  This compares favorably with the slope of $dT_c/d(d/\ell) \approx -1.3$ K implied by the experimental data shown in Fig. 2(c).

The possibility that the transition we observe is the KT transition is tempered by at least two facts.  First, a true KT transition is only expected in the unrealizable limit of zero tunneling.  However, since the estimated  tunnel splitting in our samples \cite{Spielman00} is roughly $10^6$ times smaller than the mean Coulomb energy $e^2/\epsilon\ell$, the KT transition might be replaced by a virtually indistinguishable cross-over.  Second, although recent experiments \cite{Kellogg04,Tutuc04,Wiersma04} have revealed extremely low dissipation in counterflow transport, the predicted \cite{Wen92,Yang94,Moon95} linear response superfluidity at \nt\ has so far failed to show up.  Whether this completely eliminates a KT-based explanation for the phase transition reported here is not known.

Alternatively, the phase transition to the coherent \nt\ state might be a first order one \cite{ Schliemann01,Stern02}.  Assuming this we can attempt to understand the variation of the critical temperature with effective layer separation via a free energy argument.  For example, the simplest scenario for the incoherent state at large \dl\ is two independent composite fermion (CF) liquids.  By analogy to an ordinary Fermi liquid at zero magnetic field, a free energy of the form $F_i = E_{i,0}-\alpha T^2$ might then be assumed \cite{FCF}.  For the coherent quantum Hall phase the entropy at low $T$ might plausibly be dominated by the observed \cite{Spielman01} linearly dispersing long-wavelength 2D pseudospin waves, the charged excitations being gapped out.  In this case the free energy would be $F_c = E_{c,0}-\beta T^3$.  Close to the zero temperature critical point (i.e. where $E_{i,0} = E_{c,0}$ and $(d/\ell)_c = (d/\ell)_{c,0}$) the CF entropy dominates and transition temperature would scale as $T_c \sim [(E_{i,0}-E_{c,0})/\alpha]^{1/2}$.  In order for this to agree with our experimental observation that $T_c$ scales linearly with \dl, the energy difference $E_{i,0}-E_{c,0}$ would have to scale quadratically with the ``distance'' $|(d/\ell)_{c,0} - d/\ell|$ from the zero temperature critical layer separation.  We are unaware of what mechanism might eliminate a simple linear dependence of $E_{i,0}-E_{c,0}$ on this quantity.

In summary, interlayer tunneling spectroscopy has been used to map out the phase boundary between the interlayer
coherent \nt\ bilayer quantum Hall phase and the incoherent states at larger layer separation as functions of
temperature and interlayer charge imbalance. Our data offers strong evidence that a finite temperature phase
transition is present in this system. 

We thank S. Das Sarma, A.H. MacDonald, X.G. Wen, K. Yang, and especially G. Refael for discussions,
and I.B. Spielman, G. Granger and L.A. Tracy for technical help. This work was supported via DOE grant DE-FG03-99ER45766 and NSF grant DMR-0552270.


\begin{references}
\bibitem{offset} Close to the phase boundary the readily identified tunnel resonance sits atop a small background conductance.  This background conductance has been subtracted from the $G(0)$ values discussed here.
\bibitem{KTfunc} This function, with \dl\ replaced by temperature, appears in the theory\cite{Halperin_Nelson,Kosterlitz} of disordered thin film superconductors.  We employ it here because of its singularly smooth behavior near the critical point. Refael, however, has suggested that there may be a deeper connection\cite{Refael}.
\end{references}
\end{document}